\documentstyle[12pt]{article}
\newcommand{\be}{\begin{equation}}
\newcommand{\ee}{\end{equation}}
\newcommand{\ba}{\begin{eqnarray}}
\newcommand{\ea}{\end{eqnarray}}
\newcommand{\nonu}{\nonumber}

\newcommand{\dsl}
  {\kern.06em\hbox{\raise.15ex\hbox{$/$}\kern-.56em\hbox{$\partial$}}}

\newcommand{\Dsl}{\not\!\! D}

\newcommand{\eeq}{\end{equation}}
\newcommand{\eeqarr}{\end{eqnarray}}
\newcommand{\ZZ}{{\rm \kern 0.275em Z \kern -0.92em Z}\;}
\begin{document}
\title{Uniqueness of Bogomol'nyi equations 
and 
Born-Infeld like  supersymmetric theories}
\author{H.R.~Christiansen$^a$, 
C.~N\'u\~nez$^b$  
and F.A.~Schaposnik$^b$\thanks{Investigador CICBA}
\vspace{0.15cm}
\\
{\normalsize\it
$^a$Centro Brasileiro de Pesquisas F\'\i sicas, CBPF-DCP}\\
{\normalsize\it
Rua Dr.~Xavier Sigaud 150, 22290-180 Rio de Janeiro, Brazil
\vspace{0.15cm}
}\\
$^b${\normalsize\it
Departamento de F\'\i sica, Universidad Nacional de La Plata}\\
{\normalsize\it
C.C. 67, 1900 La Plata, Argentina}}
\date{\today}
\maketitle
\begin{abstract}
We discuss Bogomol'nyi equations for  general gauge theories 
(depending on the two Maxwell invariants
$ F^{\mu \nu} F_{\mu \nu}$ and $ \tilde F^{\mu \nu} F_{\mu \nu}$)
coupled to Higgs scalars.
By  analysing their supersymmetric extension,
we explicitly show why the resulting BPS structure is insensitive
to the particular form of the gauge Lagrangian:
Maxwell, Born-Infeld
or more complicated non-polynomial Lagrangians all satisfy 
the same Bogomol'nyi equations and bounds which are dictated 
by the underlying supersymmetry algebra.
\end{abstract}


\bigskip

\newpage

Supersymmetric extensions of Born-Infeld theories
and the corresponding Bogomol'nyi equations play 
a central r\^ole in the dynamics of D-branes
and for this reason they  have
recently received a lot of attention
\cite{Tse}-\cite{GGT}. SUSY Born-Infeld theories
 were originally studied in \cite{DP}-\cite{CF} (see also \cite{H}) while
for the Bogomol'nyi equations in Born-Infeld systems, early constructions 
were reported
in \cite{NS1}-\cite{NS2}. More recent analyses of these issues 
were presented in \cite{G}-\cite{GGT}, \cite{GNSS}.

When studied \`a la Bogomol'nyi, Born-Infeld theories were coupled 
to Higgs scalars so as to reproduce the ordinary 
(i.e. Maxwell or Yang-Mills) BPS  relations \cite{NS1}-\cite{NS2}.
From the supersymmetry point of view, the usual BPS equations
arise naturally on very general grounds \cite{CM}-\cite{GGT}, 
\cite{GNSS}. As a consequence of these results, the question on whether
BPS relations are or not sensitive to the 
dynamics that one choses for the gauge field is then posed. It is the
purpose of this letter to give an answer to this question.

A first quick answer can be drawn by observing
that in the SUSY framework, one obtains Bogomol'nyi equations by 
imposing the vanishing of (half of) the supersymmetry
variations of the gaugino and higgsino fields
and these variations are formally the same for very different
Lagrangians. The dynamics associated with the Lagrangian enters however
through the equation of motion for the auxiliary field $D$
(of the gauge field supermultiplet) which appears in
the supersymmetric transformation law for the gaugino.   
It is then through 
$D$ that the form of the Lagrangian may in principle 
determine the form of the BPS relations.

What we show in this
work is precisely that supersymmetry together
with the (algebraic) equation of motion for $D$ make
the BPS relations remain unchanged irrespectively of the
choice of the gauge field Lagrangian. 

We present our analysis by considering
an  Abelian gauge theory in $d=3$ dimensions, for which the Bogomol'nyi 
equations are those originally derived in \cite{Bogo}-\cite{dVS}
for a Maxwell action; 
our arguments should hold, however, for other models
like for example the $SO(3)$ gauge theory and in other dimensionalities 
of space-time. 

 ~

Our conventions will be those of reference \cite{GNSS}.

We first consider $d=4$ dimensional Minkowski space  (with signature
$(+,-,-,-)$) and then proceed to dimensional reduction to $d=3$.
The gauge vector superfield $V$ is written, in the Wess-Zumino gauge,
\be
V = - \theta \sigma^\mu \bar \theta A_\mu + 
i \theta \theta \bar \theta \bar \lambda - 
i \bar \theta \bar \theta \theta \lambda
+ \frac{1}{2} \theta  \theta  \bar \theta  \bar \theta D 
\label{1}
\ee
Here $A_\mu$ is a vector field, $\lambda = (\lambda_\alpha)$ and 
$\bar \lambda = (\bar \lambda^{\dot \alpha})$ are two-component spinors
($\alpha, \dot \alpha = 1,2$) which can be combined to give  
a four-component Majorana fermion
and $D$ is an auxiliary field.

From $V$ a chiral superfield $W_\alpha$ can be constructed,
\begin{equation}
W_\alpha \left( y,\theta ,\bar \theta \right) =-i\lambda _\alpha +\theta
_\alpha D-\frac i2\left( \sigma ^\mu\bar \sigma ^\nu\theta \right) _\alpha
F_{\mu\nu}+\theta \theta \left( \sigma ^\mu\partial _\mu\bar 
\lambda \right) _\alpha 
\label{2}
\end{equation}
Here $\lambda ,$ $\bar \lambda ,$ $D$ and $F_{\mu\nu} = \partial_\mu A_\nu
- \partial_\nu A_\mu$   are functions of
the variable $y^\mu=x^\mu+i\theta \sigma ^\mu\bar \theta $ where
$x^\mu$ is the usual 4-vector position.
The SUSY
extension of (standard) gauge-invariant (Maxwell, Yang-Mills) theories are
precisely constructed  from $W$ by considering 
$W^2$ and its hermitian conjugate $\bar W^2$.

Now, as stressed in \cite{DP}, another superfield combination
enters into play if one wishes to construct \underline{general}
gauge invariant SUSY Lagrangians. In particular, one needs to
consider two  superfields $X$ and
$Y$ defined as
\be
X = \frac{1}{8} (D^\alpha D_\alpha W^2 + 
\bar D_{\dot \alpha}\bar D^{\dot \alpha}\bar W^2)
\label{13v}
\ee
\be
Y = -\frac{i}{16}  (D^\alpha D_\alpha W^2 - 
\bar D_{\dot \alpha}\bar D^{\dot \alpha}\bar W^2)
\label{14v}
\ee
with covariant derivatives given by
\be
D_\alpha = ~ \frac \partial {\partial \theta ^\alpha }+2i
\left( \sigma ^\mu\bar 
\theta \right) _\alpha \frac \partial {\partial y^\mu}, \ \ \ \ \ 
\bar D_{\dot\alpha} = -\frac{\partial}{\partial \bar\theta^{\dot\alpha}}  
\label{10v}
\ee
when acting on functions of $(y,\theta,\bar\theta)$ and
\be
D_{\alpha} =  \frac{\partial}{\partial \theta^\alpha}, 
\ \ \ \
\bar D_{\dot \alpha } = - \frac 
\partial {\partial \bar \theta ^{\dot \alpha }%
}-2i\left( \theta \sigma ^\mu\right) _{\dot \alpha }
\frac \partial {\partial y^{\dagger\mu}} 
\label{11v} 
\ee
on functions of $(y^\dagger,\theta,\bar\theta)$.
The only components of these superfields having purely bosonic
terms are
\ba
X |_{0} &=& - (
 D^2 - \frac{1}{2} F^{\mu \nu} F_{\mu \nu}
-i \lambda \dsl \bar \lambda -i \bar \lambda \bar{\dsl}  \lambda)
\nonu\\
X |_{\theta\bar\theta} &=& i\theta\sigma^p\bar\theta\partial_p\,
(D^2 - \frac{1}{2} F^{\mu \nu} F_{\mu \nu}
-i \lambda \dsl \bar \lambda -i \bar \lambda \bar{\dsl}  \lambda)
\nonu\\
X |_{\theta\bar\theta\theta\bar\theta} &=& \frac 1 4 
\theta\bar\theta\theta\bar\theta\,\Box\,
(D^2 - \frac{1}{2} F^{\mu \nu} F_{\mu \nu}
-i \lambda \dsl \bar \lambda -i \bar \lambda \bar{\dsl} \lambda)
\label{3}
\ea
and 
\ba
Y|_{0} &=& \frac{1}{2} (\frac{1}{2} 
F^{\mu \nu}\tilde F _{\mu \nu} +  \lambda \dsl \bar \lambda 
- \bar \lambda \bar{\dsl}  \lambda)\nonu\\
Y|_{\theta\bar \theta} &=& -\frac{i}{2}\theta\sigma^p\bar\theta
\partial_p\, (\frac{1}{2} 
F^{\mu \nu}\tilde F _{\mu \nu} +  \lambda \dsl \bar \lambda 
- \bar \lambda \bar{\dsl}  \lambda)\nonu\\
Y|_{\theta\bar \theta\theta\bar \theta} &=& \frac 1 8
\theta\bar \theta\theta\bar \theta\, \Box\,
(\frac{1}{2} 
F^{\mu \nu}\tilde F _{\mu \nu} +  \lambda \dsl \bar \lambda 
- \bar \lambda \bar{\dsl}  \lambda)
\label{4}
\ea
with $\tilde F_{\mu \nu} = 
(1/2) \varepsilon_{\mu \nu \alpha \beta} F^{\alpha \beta}$.

A third superfield combination is necessary for
constructing general gauge invariant SUSY Lagrangian.
This combination is $W^2 \bar W^2$ with 
its highest component taking the form
\be
W^2 \bar W^2\vert_{\theta \theta \bar \theta \bar \theta} =
\theta \theta \bar \theta \bar \theta \left(
(D^2 - \frac{1}{2} F_{\mu \nu}F^{\mu \nu})^2 + 
(\frac{1}{2}\tilde F_{\mu \nu}F^{\mu \nu})^2 \right)
\label{5}
\ee

Remark that again, all the
dependence of  (\ref{5})    on the curvature $F_{\mu \nu}$
and the auxiliary field $D$  is through the combination 
\be
t = \frac{1}{\beta^2} \left(D^2 - \frac{1}{2} F^{\mu \nu} F_{\mu \nu}\right)
\label{6}
\ee
and this fact will have important consequences in our discussion.
Here, in order to define a dimensionless variable $t$
we have introduced a parameter $\beta$ with the same dimensions
as $F_{\mu\nu}$ (i.e. dimensions of a mass in $d=4$). It corresponds
to the {\it absolute field} in the Born-Infeld theory \cite{B}-\cite{BI}
as will become clear below.

We are  ready to write a general $N=1$ supersymmetric Lagrangian endowed 
with gauge-invariance in terms of $X$, $Y$ and $W^2\bar W^2$ 
\begin{equation}
 L^{d=4} =
\frac{1}{4e^2}
\int\left( 
W^2 d^2\theta  + \bar W^2  d^2\bar  \theta
\right) +
\frac{1}{e^2}
\sum_{r,s,t=0}^\infty a_{rst}\int d^4\theta \left(
W^2\bar W^2\right)^r X^s Y^t  
\label{7}
\end{equation}
with $e$ the fundamental gauge coupling constant, which has been
factorized in both terms for later convenience. 

As it happens for the last component of $W^2\bar W^2$,
the first term in the r.h.s. of eq.(\ref{7}) depends on $F_{\mu \nu}$
and $D$ through the combination (\ref{6}). Indeed, the last
component in $W^2$ ($\bar W^2$) contains the term
$D^2 - \frac{1}{2} F^{\mu \nu} F_{\mu \nu} + i F_{\mu \nu}\tilde F^{\mu \nu}$
($D^2 - \frac{1}{2} F^{\mu \nu} F_{\mu \nu} -
 i F_{\mu \nu} \tilde F^{\mu \nu}$) so that the sum of $\theta$
 ($\bar \theta$ )integrals
leads to the well-known SUSY
extension of the Maxwell theory. The second term accounts
for the non-polynomial features of the
general bosonic theory to be supersymmetrized. As explained in 
\cite{DP}, supersymmetry
imposes two constraints on coefficients $a_{rst}$.
Their explicit form  will not
be relevant for our discussion. What one should retain is
that expression (\ref{7}) gives then the most general Lagrangian  
corresponding to the supersymmetric extension of a general bosonic 
Lagrangian depending on the two algebraic Maxwell invariants 
$ F^{\mu \nu} F_{\mu \nu}$ and $ \tilde F^{\mu \nu} F_{\mu \nu}$.
 
As stated above, our actual interest is focused on a $d=3$, 
$N=2$ supersymmetric 
theory which can be obtained from Lagrangian
(\ref{7}) by dimensional reduction.
The standard procedure for dimensional reduction, 
say in the $x_3$ spatial coordinate, 
implies identifying $A_3$ with a scalar field $N$. 
Now, it can be shown that without including
a Chern-Simons term, the bosonic part of the Lagrangian (\ref{7}) 
can only yield electrically neutral configurations, so that as
long as one looks for self-dual equations associated with (static) 
vortices,  the $A_0$ field (as well as the $N$ field)
can be put to zero  and so we will do from here on 
(the case $N \ne 0$ can be equally treated without
additional complications). 
So far,  without the addition of a Chern-Simons term, 
no electrically charged vortices 
exist and then the most general gauge field configurations are
pure magnetic  Nielsen-Olesen type soliton solutions \cite{dVS0}. 
This implies that no $d=3$ version of the $\tilde F_{\mu \nu} F^{\mu \nu}$
functional are available and that we can simply identify the field strength 
with the magnetic field $B$  by
\be 
\frac{1}{2} F_{\mu \nu} F^{\mu \nu} = B^2
\label{8}
\ee
with
\be
B = \frac{1}{2} \varepsilon_{jk}F^{jk}  ~ ~ ~  i,j=1,2
\label{9}
\ee
Once the dimensional reduction is carried on, one ends with the
$d=3$ version of the SUSY Lagrangian given in eq.(\ref{7}). 
As it is well known, supersymmetry can be extended
from $N=1$ to $N=2$ in this process. 

From what we have seen, the gauge field  dependent 
terms in the $bosonic$ part of this $N=2$ supersymmetric Lagrangian 
can be compactly written in the form
\be
L_{A}[A_\mu,D] = \frac{1}{e^2}\sum_{n=0}^{\infty} c_n  
t^n
\label{10}
\ee
where $t$ (defined in (\ref{6})) now reads
\be
t = \frac{D^2 - B^2}{\beta^2}
\label{11}
\ee
and $c_n$ are some coefficients which can be computed in terms of
the $a_{rst}$'s.

Concerning the Higgs field sector, 
in $d=4$ dimensions the coupling between the scalar
Higgs field $\phi$ and the gauge field $A_\mu$ arises from the superfield
interaction term
\be
L_{A-\phi}^{d=4} = \Phi \exp(V) \Phi^*
\label{12}
\ee
where $\Phi$ is a chiral scalar superfield containing a Higgs field
$\phi$, a higgsino $\psi$ and an auxiliary field $F$.  One
can easily see that the part of $L_{A-\phi}$ containing the auxiliary
field $D$ is  \cite{ENS}
\be
L_{A-\phi}^{d=4}|_D = \frac{1}{2}  D \vert \phi \vert^2
\label{13}
\ee
On the other hand, gauge symmetry breaking  can be achieved {\it \`a la} 
Fayet-Iliopulos  so that the complete $D$ dependence of 
the supersymmetric Lagrangian arising from the Higgs coupling to $A_\mu$
and $D$ is given by
\be
L^{d=4}_D [A,\phi,D] \equiv L^{d=4}_{A-\phi}|_D + L^{d=4}_{FI} = 
  \frac{1}{2} D(|\phi|^2 - \xi^2)
\label{14}
\ee
where $\xi$ is a real constant.
This Lagrangian remains unchanged after dimensional reduction
so that  we can write  the  $D$ dependent 
terms of the $d=3$  bosonic part of the Lagrangian as
\be
L_D^{total}[A,\phi,D] = \frac{1}{e^2} \sum_{n=0}^{\infty} c_n  
\left(\frac{1}{\beta^2}( D^2 - B^2)\right)\!^n  +
 \frac{1}{2} D(|\phi|^2 - \xi^2)
\label{15}
\ee
In $d=3$ space-time, dimensions of parameters and fields
are $[\beta]= m^2$, $[e] = m^{\frac{1}{2}}$, $[\xi] = m^{\frac{1}{2}}$,
$[A_\mu] = m$, $[D] = m^2$ and $[\phi] = m^{\frac{1}{2}}$. Then,
for dimensional reasons, one can infer that coefficients $c_n's$ can
be written in the form
\be
c_n = \beta^2 \lambda_n
\label{17}
\ee
where $\lambda_n$ are dimensionless coefficients.
It should be noted that the choice
of $\lambda_1 = -1/2$, $\lambda_n = 0$ for $n \ne 1$
corresponds to the usual
value of the Maxwell term while the choice $\lambda_1 = -1/2$,
$\lambda_2 = 1/8$, $\lambda_3  = 1/32$, $\ldots$,
gives a Born-Infeld Lagrangian for the gauge field.

We can now obtain the equation of motion for  
$D$ so as to eliminate the auxiliary field from the physical spectrum
\be
\sum_{n=0}^{\infty}  \frac{2n}{e^2} \lambda_n 
\left(\frac{1}{\beta^2}( D^2 - B^2)\right)\!^{n - 1} {D} +
 \frac{1}{2}(|\phi|^2 - \xi^2) = 0
\label{16}
\ee
One can easily see that the only nontrivial solution to eq.(\ref{16}) 
takes the form
\begin{eqnarray}
D & = & - \frac{e^2}{4\lambda_1}(|\phi|^2 - \xi^2 \nonumber)\\
B & = & \pm D
\label{17v}
\end{eqnarray}
These two equations can be readily combined into one  which is nothing
but the well-honored Bogomol'nyi equation for the magnetic field
of the Nielsen-Olesen vortices 
\be
B = \mp \frac{e^2}{4\lambda_1}(|\phi|^2 - \xi^2)
\label{18}
\ee
This shows that the Bogomol'nyi gauge
field equation for vortex configurations is \underline{independent} 
of the particular
form of the gauge field Lagrangian one chooses since
we have proven formula (\ref{18}) for the general 
supersymmetric Lagrangian (\ref{7})+(\ref{12}). 

Let us now analyse the $N=2$ supersymmetry 
transformations  leaving invariant the three dimensional
Born-Infeld SUSY theory. 
We shall not write the complete set of transformations but just
those which are relevant for the discussion of
Bogomol'nyi equations, namely those for the higgsino and
gaugino (which we call $\psi$ and $\Sigma$): 
\be 
\delta_\epsilon \psi = -i \Dsl \phi \epsilon  
= \left( \begin{array}{cc} 0 & D_1 + i D_2\\
	                   D_1 - i D_2 & 0 \end{array} \right)
\left( \begin{array}{c} \epsilon_+\\
                        \epsilon_- \end{array} \right)                     
\label{19}
\ee
\be
\delta_\epsilon \Sigma = (\frac{1}{2}\varepsilon_{\mu\nu\alpha}
F^{\mu\nu}\gamma^\alpha + D) \epsilon =
 \left( \begin{array}{cc} \frac{1}{2} \varepsilon_{ij}F^{ij} + D & 0\\
	                  0 & \frac{1}{2} \varepsilon_{ij}F^{ij} - D 
\end{array} \right)
\left( \begin{array}{c} \epsilon_+\\
                        \epsilon_- \end{array} \right)  
\label{20}
\ee
where we call $\epsilon$ the Dirac fermion 
transformation parameter
(we have already made $N = A_0 = 0$ and considered the static case).

As it is well-known, making zero half of the SUSY variations 
associated with the higgsino and gaugino fields, 
one gets the Bogomol'nyi equations. For
instance, by demanding that those generated by
$\varepsilon_+$ be zero, one gets the following
 self-dual equation from the higgsino's
variation
\be
\delta_{\epsilon_+} \psi = 0 \to D_1 \phi = i D_2 \phi
\label{21}
\ee
One should note that this transformation law just depends on the
way the parallel displacement is defined in terms of the gauge
connection and not on the explicit form of the gauge field action. One
can then understand why eq.(\ref{21}) is completely independent of the
particular form the gauge field action is chosen, at least for minimally 
coupled gauge theories \footnote{For an analysis of Bogomoln'yi equations 
in non-minimally coupled gauge theories, see ref.\cite{cbpf}.}. 
Regarding the equation derived from the gaugino transformation,
\be
\delta_{\epsilon_+} \Sigma = \frac{1}{2} \varepsilon_{ij}F^{ij} + D = 0
\label{22}
\ee
it could, in principle, depend on the particular Lagrangian chosen
through the $D$ term. However, as we have seen (eq.(\ref{17v})), 
the solution to the  equation of motion for $D$ takes the same simple
form for any gauge field Lagrangian since
$D$ always enters through the combination $D^2 - B^2$.

This feature can be also checked by analyzing the two supercharges which 
can be obtained
following the usual Noether construction. As it has been shown in
\cite{GNSS} for the Born-Infeld case,   
supercharges $Q$ and $\bar Q$ can be always put in the form
\ba 
\bar Q &=& i \int d^2x\ \Sigma^\dagger\ {\cal H}[B,D]\ (\gamma^0 B +D) +
\frac{i}{2} \int d^2x\ \psi^\dagger\ {\Dsl \phi}\nonumber \\
Q &=& -i\int d^2x\  (B +\gamma^0 D)\ {\cal H}[B,D]\ \Sigma - 
\frac{i}{2} \int d^2x\ \gamma^0 ({\Dsl \phi})^\dagger \psi
\label{24}
\ea
with ${\cal H}$ some real functional of $D$ and $B$
which can be computed
order by order in $1/\beta^2$. Furthermore, eqs.(\ref{24}) also hold
when one considers not just SUSY Born-Infeld theory but
the general Lagrangian, viz. eq.(\ref{7}).  
Only the actual form of ${\cal H}$ will
change, depending on the different sets of possible $a's$ coefficients.
What one can easily see is that the following formula holds
\be
{\cal H} = {\cal H}_{Maxwell}+ \sum_{n=1}^{\infty}\frac{1}{\beta^{2n}} 
{\cal H}_n[B,D]
\label{gene}
\ee
with 
\be
{\cal H}_{Maxwell} = 1
\label{mu}
\ee
\be
\left.{\cal H}_n [B,D]\right\vert_{B^2 = D^2} = 0
\label{que}
\ee

It is clear that condition $\bar Q |phys\rangle = 0$ is 
satisfied whenever $(B  +\gamma^0 D)\epsilon =0$ 
and  $ {\Dsl\phi} \epsilon  = 0$, 
independently of the precise form the functional ${\cal H}$ takes. 
Choosing just the upper component of the transformation parameter,  
$\epsilon_+$, yields again 
the two Bogomol'nyi equations (\ref{18}),(\ref{21}). Of course, this
is to be expected since both $(B  + \gamma^0 D)\epsilon$ and 
$ \Dsl \phi \epsilon$,
appearing in (\ref{24}), provide the transformation laws of gaugino
and higgsino respectively.

Concerning the supercharge algebra,  when the Bogomol'nyi 
equation $B = \pm D$ is used, only the
Maxwell part of ${\cal H}$ survives, this showing again why
the BPS structure is not sensitive to the particular form of the 
gauge field Lagrangian.

 ~

In conclusion, we have analysed the most general Lagrangian  
corresponding to the supersymmetric extension of a general 
bosonic Lagrangian depending on the two algebraic Maxwell invariants  
$ F^{\mu \nu} F_{\mu \nu}$ and $ \tilde F^{\mu \nu} F_{\mu \nu}$.
This general Lagrangian includes, for a particular choice of
coefficients, the Born-Infeld supersymmetric
Lagrangian, and also an infinite class of Lagrangians having causal
propagation \cite{DP}. We have shown why the Bogomol'nyi relations
associated with the bosonic sector remain unchanged in spite of
of the actual form of the gauge field Lagrangian: Maxwell, Born-Infeld
or more complicated non-polynomial Lagrangians all have the same
BPS structure.

Finally, we  note that a similar analysis 
could be in principle undertaken for the case of non-abelian
gauge theories. There is in this respect some ambiguity about the 
way one should define the trace structure of the action. For the 
Born-Infeld case, it is pointed
out in \cite{Bre} that the symmetrised trace defined in \cite{Tse2}
seems to be singled out by BPS considerations with respect to other 
definitions, this suggesting the existence of a supersymmetric
extension of such an action  but not
in those with other trace structure. We think precisely that a
combined analysis of the general supersymmetry transformations 
and the equation of motion for the auxiliary field 
analogous to that presented here
will show what kind of actions admit BPS relations.
 We hope to report on this issue in a separate work.

 ~  

\underline{Acknowledgements}: H.R.C. is partially supported by 
FAPERJ, Rio de Janeiro, Brazil, and C.N. by CONICET, Argentina.
F.A.S. is partially supported by CICBA, CONICET, Fundaci\'on Antorchas,
Argentina and a Commission of the European Communities
contract No:C11*-CT93-0315.


\end{document}